\documentclass[prb,preprint]{revtex4-1}
\usepackage{amsmath}  % needed for \tfrac, \bmatrix, etc.
\usepackage{amsfonts} % needed for bold Greek, Fraktur, and blackboard bold
\usepackage{graphicx} % needed for figures
\begin{document}
% Be sure to use the \title, \author, \affiliation, and \abstract macros
% to format your title page.  Don't use lower-level macros to  manually
% adjust the fonts and centering.
\title{A Simple Derivation of Lorentz Self-Force}
% In a long title you can use \\ to force a line break at a certain location.
\author{Asrarul Haque}
\email{ahaque@hyderabad.bits-pilani.ac.in} % optional
\altaffiliation[B209, Department of Physics ]{BITS Pilani Hyderabad
campus, Jawahar Nagar, Shameerpet Mandal, Hyderabad-500078, AP,
India.} % optional second address
% If there were a second author at the same address, we would put another
% \author{} statement here.  Don't combine multiple authors in a single
% \author statement.
\affiliation{Department of Physics, BITS Pilani Hyderabad Campus,
 Hyderabad-500078, AP, India.}
% Please provide a full mailing address here.
% See the REVTeX documentation for more examples of author and affiliation lists.
\date{\today}
\begin{abstract}
%We address the issue of causality in classical physics.
We derive the Lorentz self-force for a charged particle in
arbitrary non-relativistic motion via averaging the retarded fields. %on the surface of sufficiently small spherical shell
 The derivation is simple and at the same time pedagogically
accessible. We obtain the radiation reaction for a charged particle
moving in a circle. We pin down the underlying concept of mass
renormalization.
\end{abstract}
% AJP requires an abstract for all regular article submissions.
% Abstracts are optional for submissions to the "Notes and Discussions" section.
\maketitle % title page is now complete
\section{Introduction}
%\subsection{Average Field}
%The field due to a point charge on itself is called the self field.
The electromagnetic field goes to infinity at the position of a
point charge. The electrostatic field at the position of the charged
particle
\begin{equation}
\vec E(\vec r) = \frac{1} {{4\pi \varepsilon _0 }}\frac{q} {{r^3
}}\vec r\xrightarrow{{r \to 0}}\infty
\end{equation}
and the self-energy of the point charge in the rest frame of the
charged particle
\begin{equation}
U = \frac{{\varepsilon _0 }}{2}\int {E^2 } d^3 r =\frac{1}{{32\pi ^2
\varepsilon _0 }}\int\limits_0^\pi  {\sin \theta d\theta
\int\limits_0^{2\pi } {d\phi \int\limits_0^\infty {\frac{{q^2
}}{{r^2 }}dr} } }  =  - \frac{{q^2 }}{{8\pi\varepsilon _0 }}\left[
{\frac{1}{r}} \right]_{r = 0}^{r = \infty }  \to \infty
\end{equation}
turn out divergent. The concept of the point-like (structureless or
dimensionless object) charge may be an idealization. The measurement
of the anomalous magnetic moment of an electron based on the quantum
field theoretic calculations leads to the following upper bound on
the size of the electron \cite{kin} $l:~l\le 10^{-17}cm $. Thus, a
charged particle
might be considered as an extended object with a finite size.\\
In order to circumvent the problem of divergence of the field at the
point charge, it is plausible either: (1) to consider the averaged
value of the field in the suitably small closed region surrounding
the point charge as the value of the field under consideration at
the position of the point charge or (2) to think a charged particle
as an extended object having small
dimension with a charge distribution. Our derivation of the self-force is based on the former consideration.\\
If an extended charged particle moves with non-uniform velocity, the
charge elements comprising such charge distribution, begin to exert
forces on one another. However, these forces do not cancel out due
to retardation giving rise to a net force known as the self-force.
Thus, the radiating extended charged particle experiences a
self-force which acts on the charge particle itself. The Lorentz
self-force \cite[p.~753]{jack1} arising due to a point charge
conceived as a uniformly charged spherical shell of radius $s$ is
given by
 \begin{equation}
\vec F_{self}  =  - \frac{2}{3}\frac{{q^2 }}{{4\pi  \varepsilon_0
c^2s }}\dot {\vec v}(t) + \frac{2}{3}\frac{{q^2 }}{{4\pi
\varepsilon_0 c^3 }}\ddot{\vec v}(t)+ O(s)~~\textup{with}~~|\vec s|=
s \label {ccc}\end{equation} where,
\begin{itemize}
\item the quantity $\frac{2}{3}\frac{{q^2 }}{{4\pi \varepsilon
_0 c^2s }}$ in the first term stands for electromagnetic mass and
becomes
divergent %since $ m_{em} \equiv
%\frac{2}{3}\frac{{q^2}}{{4\pi \epsilon _0 c^2s }}\to \infty$
as $s\to 0^+$,
\item the second term represents the radiation reaction and is independent of
the dimension of the charge distribution and
\item the third term corresponds to the first finite size correction and is proportional to the
radius of the shell $s$.
\end{itemize}
The usual method \cite{jack1} of computing the self-force
(\ref{ccc}) involves rather cumbersome calculation. Boyer \cite{boy}
has obtained the expression for the self-force using the averaged
value of the retarded field for a charged particle in uniform
circular motion. Boyer's derivation of the self-force for charged
particle in uniform circular motion involves an unsophisticated
calculation.\\
In this article, we derive the expression for the self-force for a
point charge in arbitrary non-relativistic motion by averaging the
retarded field in rather neat and sophisticated way. Our derivation
of the self-force unlike Boyer's derivation pertaining to the
specific context, leads to the general (non-relativistic) expression
for the self-force. In the following section, we shall define the
self-force in terms of the averaged retarded field over the surface
of the spherical shell.
\section{The Self-Force}
We shall be considering for our purpose, an average field on the
surface of a spherical shell of radius $s$ due to a point charge $q$
sitting at the center of the shell. An average field \cite{cla} over
the surface of a spherical shell of radius $s$ is defined by
\begin{equation}
\overline{\vec E}_q(t)  =
%\mathop{Lim}\limits_{s \to 0^+}
\frac{1}{{4\pi s^2 }}\int\limits_{\sum} {dA} \vec E(\vec r,t)
\end{equation}
where $\sum$ is the surface of the spherical shell of radius $s$.
 We now define the self-force as
\begin{equation}
\vec F_{Self} = q\mathop {Lim}\limits_{s \to 0^+} \overline{\vec
E}_q(t)  = q\vec E_{Self}
\end{equation} where the field ${\vec E}(\vec r,t)$ depends upon the
position and motion of the charge particle at the retarded time. The
field due to an accelerated charged particle is the sum of the
\emph{velocity fields} as well as the \emph{acceleration fields}:
\begin{equation} \vec E(\vec r,t)  = \vec E^{Vel}(\vec r,t) + \vec E^{Acc}(\vec r,t).  \end{equation}
 The average field contribution from the \emph{velocity fields}
 (for $v<<c$) $\overline{ \vec E_{q}^{Vel}}(t)$ on
the surface of a spherical shell due to a charged particle at its
center gets filtered out of $\overline{\vec E}_q(t)$ because for
each spatial point ( say ($s_x,s_y,s_z$)) on the surface there
exists a corresponding point
%reflection symmetry(The reason for considering the surface of a spherical shell,
($-s_x,-s_y,-s_z$) on the surface.\\
However, the average field contribution from the \emph{acceleration
fields} $\overline{ \vec E_{q}^{Acc}}(t)$ over the surface of a
spherical shell due to a charged particle situated at the center of
the shell turns out nonzero because the \emph{acceleration fields}
involve a term $\vec s(\vec s.\vec a)$ which is even in the vector
$\vec s$.
% and will therefore
%yield nonzero value.
Moreover, $\overline{ \vec E_{q}^{Acc}}(t)$ leads to radiation
reaction as well as a term associated with the electromagnetic
contribution to mass. We shall now obtain the derivation of the
self-force in the following section.
\section{Radiation Reaction From a Point Charge in Arbitrary Non-relativistic Motion}
In order to obtain the expression for the self-force, we begin with
%\subsection{Definition of Self Field}
%\subsection{Calculation of Radiation Reaction}
the field \cite[p.~664]{jack1} due to a point charge located at
$\vec r_q(t)$ at time $t$ in an arbitrary motion which is given by
\begin{equation}
\vec E(\vec r,t) = \left[ { \frac{q {\left( {\vec s - \frac{{\vec
v}}{c}s} \right)\left( {1 - \frac{{v^2 }}{{c^2 }}} \right) }}{{4\pi
\varepsilon _0 \left| {\vec s - \frac{{\vec v}}{c}s} \right|^3}} +
\frac{q\left ( \vec s \times \left( {\vec s - \frac{{\vec v}}{c}s}
\right) \times \frac{{\vec a}}{{c^2 }}\right)}{{4\pi \varepsilon _0
\left| {\vec s - \frac{{\vec v}}{c}s} \right|^3}} } \right]_{t =
t_{{\mathop{\rm Re}\nolimits} t} }
\end{equation}
where,
\begin{equation}
 \vec s = \vec r - \vec
r_q(t_{Ret}) = (x - x_q(t_{Ret}))\hat i + (y - y_q(t_{Ret}))\hat j +
(z - z_q(t_{Ret}))\hat k,
 \end{equation}
and $t_{Ret}=t-s/c$. The quantities $\vec s,~\vec v$ and $\vec a$ in
the square brackets are evaluated at the retarded time $t_{Ret}$.
The first term represents the \emph{velocity fields} $\vec
E^{Vel}(\vec r,t)$
 and is the independent of the acceleration of the point charge. The second term represents
 \emph{acceleration fields} $\vec E^{Acc}(\vec r,t)$ and exists only
when $\vec a =\frac{{d\vec v}}{{dt}} \ne 0$. Thus, for $\vec a=0$,
there will be no radiation. For sufficiently small speed of the
point charge ( in the limit $v/c\to 0$ ) the field takes the form
\begin{eqnarray}
\vec E (\vec r,t) &=& \frac{q}{{4\pi \varepsilon _0 }}\frac{\vec s
}{{\left| {\vec s} \right|^3 }}+ \frac{q}{{4\pi \varepsilon _0
}}\frac{1}{{\left| {\vec s} \right|^3 }}\left[ {\vec s \times \left(
{\vec s \times \frac{{\vec a}}{{c^2 }}} \right)}
\right]\nonumber\\
&=&\frac{q}{{4\pi \varepsilon _0 }}\frac{\vec s }{{\left| {\vec s}
\right|^3 }}+ \frac{q}{{4\pi \varepsilon _0 }}\frac{1}{{c^2
s^3}}\left[ {\vec s(\vec s.\vec a) - s^2\vec a} \right]
\end{eqnarray}
\begin{figure}[hbtp!]
\begin{center}
    \includegraphics[bb = 100 300 370 550, scale=0.7,angle=0]{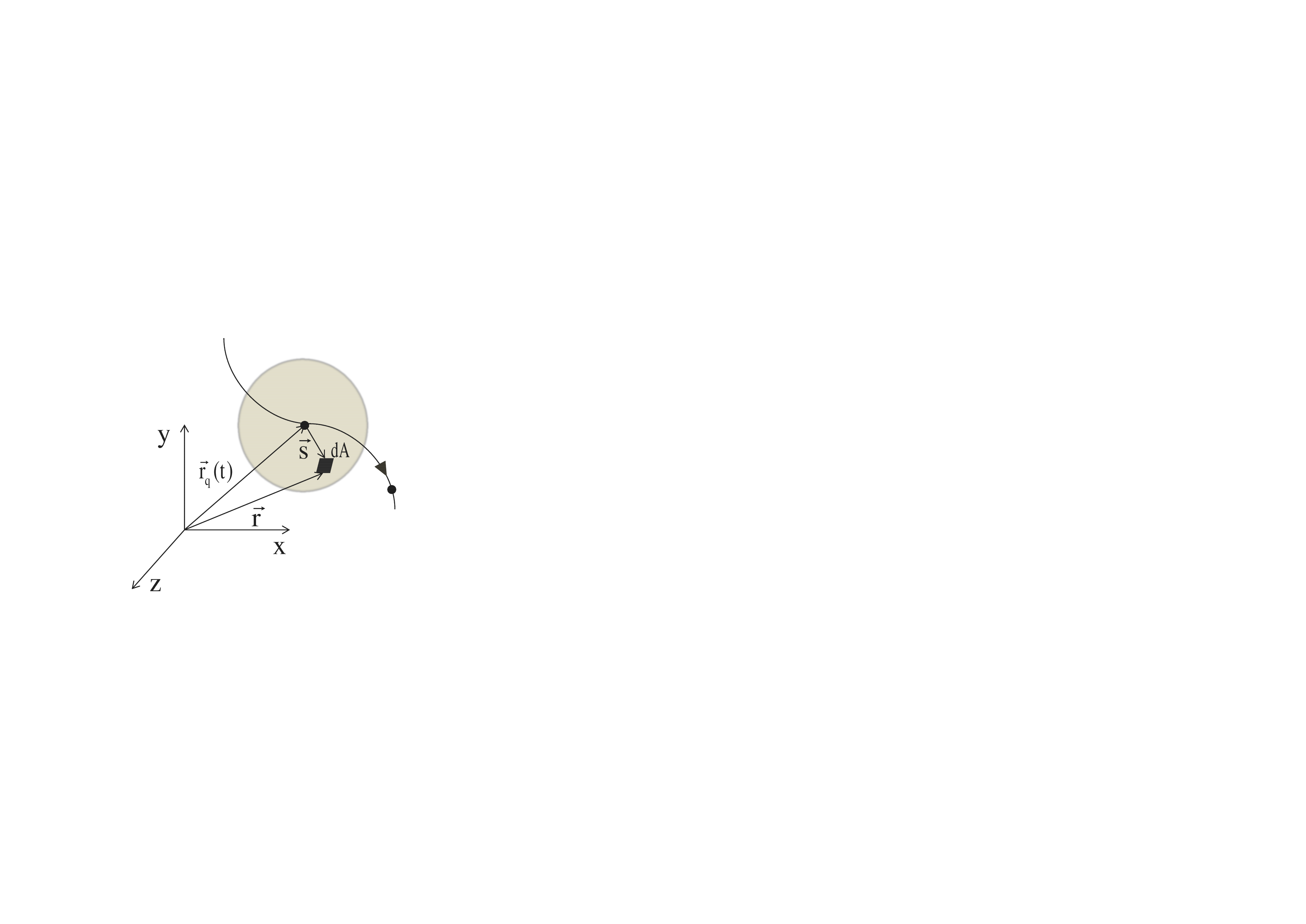}
    \caption{A charged particle in arbitrary motion.}
\label{casa}
\end{center}
\end{figure}
%Where,
%\begin{equation}
% \vec s = \vec r - \vec r'(t_{Ret}) = (x - x'(t_{Ret}))\hat i + (y - y'(t_{Ret}))\hat j + (z - z'(t_{Ret}))\hat k
% \end{equation}
%%%%%%%%%%%%%%%%%%%%%%%%%%%%%%%
The velocity and corresponding acceleration of the point charge are
given by
\begin{eqnarray}
\vec v &=& \dot{\vec r}_q(t_{Ret}) = \dot x_q(t_{Ret})\hat i + \dot
y_q(t_{Ret})\hat j + \dot
z_q(t_{Ret})\hat k \nonumber\\
\vec a &=& \ddot{\vec r}_q(t_{Ret}) = \ddot x_q(t_{Ret})\hat i +
\ddot y_q(t_{Ret})\hat j + \ddot z_q(t_{Ret})\hat k
\end{eqnarray}
Now, the average field over the surface of sphere (as shown in the
FIG.
\ref{casa})  is %$\vec E (\vec r,t)$
%\[
%\vec E_{Rad}  = \frac{q}{{4\pi  \in _0 }}\frac{1}{{c^2 s}}\left[
%{\hat s(\hat s.\vec a) - \vec a} \right]
%\]
%%%%%%%%%%%%%%%%%%%%%%%%%%%%%%%
\begin{eqnarray}
 \overline{\vec E}_q(t)  &=& \frac{1}{{4\pi s^2 }} \iint { d\theta d\phi s^2 \sin \theta}
  \vec E (\vec r,t) \nonumber\\
  &=& \frac{q}{{4\pi  \varepsilon _0 }}\frac{1}{{c^2 }}\frac{1}{{4\pi s^3 }} \iint {
   d\theta d\phi \sin \theta } \left[ {c^2\vec s + \vec s(\vec s.\vec a) - s^2\vec a}
   \right]
 \end{eqnarray}
where $\vec s(\vec s.\vec a)$ may be expressed as
%%%%%%%%%%%%%%%%%%%
\begin{eqnarray}
 \vec s(\vec s.\vec a) &=& \left[ {\vec s(s_x a_x  + s_y a_y  + s_z a_z )} \right]\nonumber \\
  &=&  (s_x^2 a_x  + s_x s_y a_y  + s_x s_z a_z )\hat i \nonumber\\
  &+&  (s_y s_x a_x  + s_y^2 a_y  + s_y s_z a_z )\hat j \nonumber\\
  &+& (s_z s_x a_x  + s_z s_y a_y  + s_z^2 a_z )\hat k.
\end{eqnarray}
The vector $\vec s$ in the spherical polar coordinates is given by
\begin{equation}
 \vec s = (s_x,s_y,s_z)= (s\sin \theta \cos \phi, s\sin \theta \sin\phi, s\cos
 \theta)
\end{equation}
%We have
%\begin{equation}
%\int {s_i s_j d\theta d\phi \sin } \theta  = \frac{{4\pi }}{3}s^2
%\delta _{ij} \end{equation}
% Now,
%%%%%%%%%%%%%%%%%%%%%%%%%%%%%%%
%\begin{eqnarray}
%&&\iint {s_x } \sin \theta d\theta d\phi = s \int\limits_0^\pi
% {\sin^2 \theta d\theta } \int\limits_0^{2\pi } {\cos \phi d\phi }
%  =0~~\textup{and}\nonumber\\
% &&\iint {s_x^2 } \sin \theta d\theta d\phi = s^2 \int\limits_0^\pi
%  {\sin ^3 \theta d\theta } \int\limits_0^{2\pi } {\cos ^2 \phi d\phi }
% s^2 \int\limits_{ - 1}^{ + 1} {(1 - \cos ^2 \theta )d\cos \theta } \int\limits_0^{2\pi }
%{\frac{{(1 + \cos 2\phi )}}{2}d\phi }
%  = \frac{{4\pi }}{3}s^2.
% \end{eqnarray}
We can show that
\begin{eqnarray}
\iint {s_i \sin \theta  d\theta d\phi }  &=& 0\\
 \iint {s_i
s_j\sin\theta d\theta d\phi}   &=& \frac{{4\pi }}{3}s^2 \delta _{ij}
\end{eqnarray}
%%%%%%%%%%%%%%%%%%%%%%%%%%%%%%%
where $\delta _{ij}=1$ for $i=j$ and $0$ otherwise with $i,j =
x,y,z$. The average field now becomes
%%%%%%%%%%%%%%%%%%%%%%%%%%%%%%%
\begin{equation}
\overline{\vec E}_q(t)   = \frac{q}{{4\pi  \varepsilon _0 c^2
s}}\left[ { \frac{1}{3}\vec a(t_{Ret} ) - \vec a(t_{Ret} )} \right]
= \frac{q}{{4\pi \varepsilon _0 c^2 s}}\left[ { - \frac{2}{3}\vec
a(t_{Ret} )} \right]. \end{equation}
%The retarded time $t_r$  is
%\begin{equation}
%t_r  = t - \frac{s}{c}
%\end{equation}
We note that the velocity fields contribution to $\overline{\vec
E}_q(t) $ vanishes for a charged particle in arbitrary
non-relativistic motion. In fact, the vanishing contribution of the
velocity fields apparently
 brings out the error in Boyer's \cite{boy}
calculation for the contribution of the velocity fields. In the
limit $s \to 0^+$, we get
\begin{equation}
 \vec a(t_{Ret}) = \vec a(t - s/c) = \vec a(t) - \frac{s}{c}\dot {\vec a}(t) + O(s^2 )
\end{equation}
Now,
\begin{equation}
\overline{\vec E}_q(t)
  =  - \frac{2}{3}\frac{q}{{4\pi  \varepsilon  _0 c^2 }}\frac{{\vec a(t)}}{s}
   + \frac{2}{3}\frac{q}{{4\pi  \varepsilon _0 c^3 }}\dot{ \vec a}(t) + O(s )
\end{equation}
Thus, the self-field $\vec E_{Self}$ yields
\begin{eqnarray}
\vec E_{Self}&=& \mathop {Lim}\limits_{s \to 0^+}\overline{\vec E}_q(t)  \nonumber \\%  = \frac{q}{{4\pi  \varepsilon  _0 c^2 s}}\left[ { - \frac{2}{3}\vec a(t) + \frac{2}{3}\frac{s}{c}\dot {\vec a}(t)} \right]+ O(s ) )\nonumber \\
  &=&  - \mathop {Lim}\limits_{s \to 0^+}\left(\frac{2}{3}\frac{q}{{4\pi  \varepsilon  _0 c^2s }}
  \right)\vec a(t) + \frac{2}{3}\frac{q}{{4\pi  \varepsilon _0 c^3 }}\dot{ \vec a}(t)% + O(s )
 \end{eqnarray}
Now, the self-force for the point charge limit is given as:
\begin{equation}
\vec F_{Self}  = q\vec E_{Self}  = - \mathop {Lim}\limits_{s \to
0^+}\left(\frac{2}{3}\frac{q^2}{{4\pi  \varepsilon_0 c^2s
}}\right)\vec a(t) + \frac{2}{3}\frac{q^2}{{4\pi  \varepsilon_0 c^3
}}\dot {\vec a}(t)%+ O(s)
\end{equation}
which is same as equation (\ref{ccc}) in the limit  $s \to 0^+$. The
self-force
 can be expressed in terms of $r_q$ and $\theta_q$ variables as
\begin{eqnarray}
\vec F_{Self}  &=&
   - \frac{2}{3}\frac{q}{{4\pi \varepsilon _0 c^2 s}}\left[ {(\ddot r_q - r_q\dot {\theta}_q ^2 )\hat r_q +
(r_q\ddot {\theta}_q  + 2\dot r_q\dot {\theta}_q )\hat {\theta}_q } \right] \nonumber\\
  &+& \frac{2}{3}\frac{q}{{4\pi \varepsilon _0 c^3 }}\left[ {( \dddot r_q- 3\dot r_q\dot {\theta}_q ^2
 - 3r_q\dot {\theta}_q \ddot {\theta}_q )\hat r_q + (r_q\dddot {\theta}_q + 3\dot r_q\ddot {\theta}_q
 + 3\ddot r_q\dot {\theta}_q
- r_q\dot {\theta}_q ^3 )\hat {\theta}_q } \right]
\end{eqnarray}We
shall now study the following illustrative examples pertaining to
radiation reaction. \subsection{Radiation Reaction From a Charged
Particle Executing Simple Harmonic Motion} Consider a charged
particle $q$ of mass $m$ executing simple harmonic motion along
X-axis with frequency $\omega$, its displacement from equilibrium is
 \[x(t) = x_0\sin\omega t \]
and its acceleration is
 \[a(t) =\ddot x = -x_0\omega^2 \sin\omega t.\]
 The charged particle having nonzero acceleration will radiate and will therefore experience
 the self-force (Please see the equation (\ref{sine}) ) given by
\begin{eqnarray}
F^{Oscillator}_{Self} &=& \mathop {Lim}\limits_{s \to 0^ +  } \left(
{ - \frac{2}{3} \frac{{q^2 }}{{4\pi \varepsilon _0 c^2 s}}}
\right)\left[ { - x_0 \omega ^2
\sin \omega \left( {t - \frac{s}{c}} \right)} \right]\nonumber \\
 & =& \mathop {Lim}\limits_{s \to 0^ +  } \frac{2}{3}\frac{{q^2 }}{{4\pi
 \varepsilon _0 c^2 s}}x_0 \omega ^2 \sin \omega t - \frac{2}{3}
 \frac{{q^2 }}{{4\pi \varepsilon_0 c^3 }}x_0 \omega^3 \cos \omega t
 \end{eqnarray}
Thus, the self-force is the sum of the finite piece $(- \frac{2}{3}
 \frac{{q^2 }}{{4\pi \varepsilon_0 c^3 }}x_0 \omega^3 \cos \omega t)$ and
the divergent piece $(\mathop {Lim}\limits_{s \to 0^ +  }
\frac{2}{3}\frac{{q^2 }}{{4\pi
 \varepsilon _0 c^2 s}}x_0 \omega ^2 \sin \omega t)$.
\subsection{Radiation Reaction From a Charged Particle Moving in a
Circle } 
\begin{figure}[hbtp!]
\begin{center}
    \includegraphics[bb = 120 230 500 550,
    scale=0.6,angle=0]{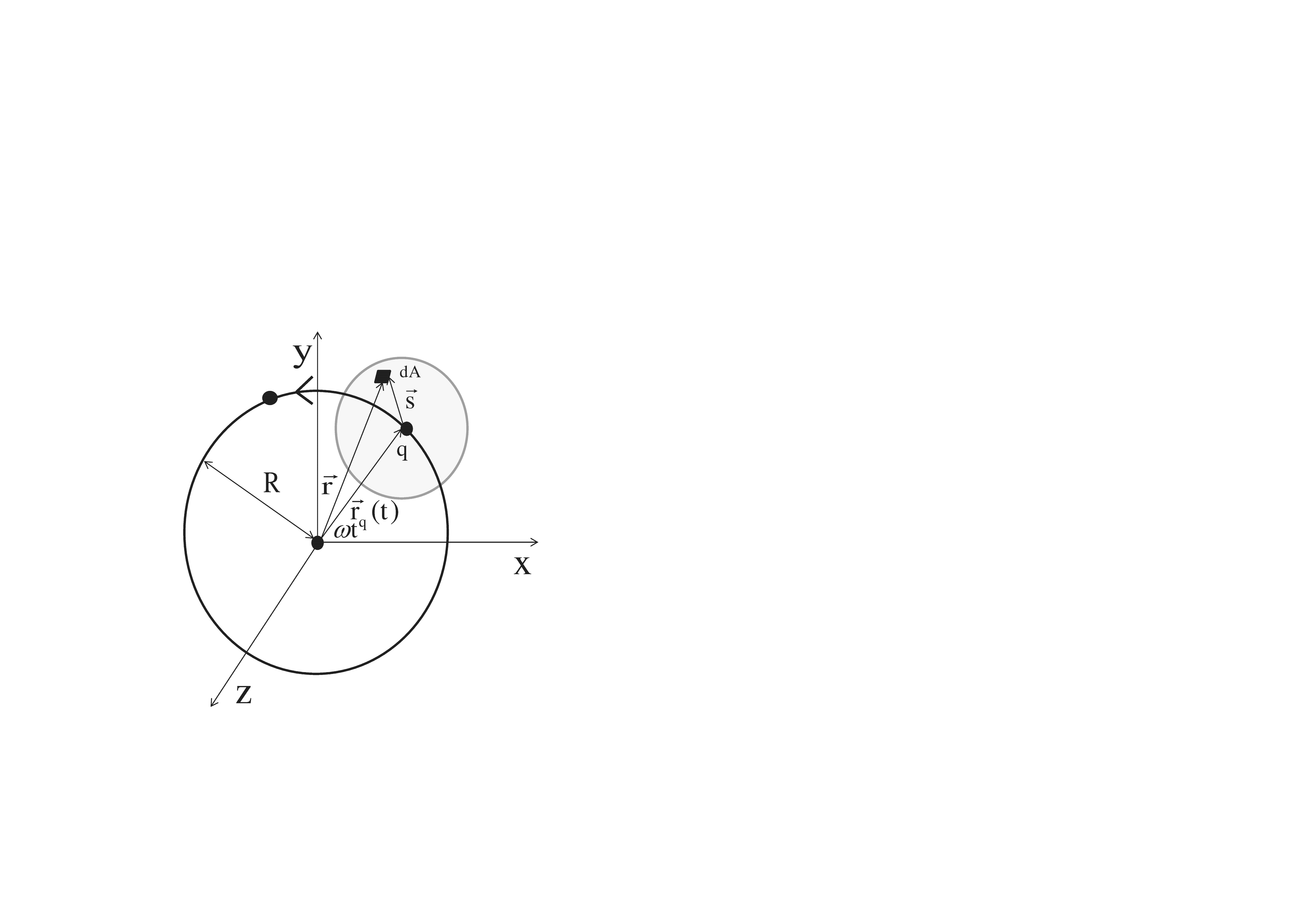}
       \caption{A charged particle moving in a circle of radius $R$.}
\label{casa9}
\end{center}
\end{figure}

Suppose a charged particle is moving in a circle of radius
$R$ with uniform angular speed $\omega$ (as shown in the FIG.
\ref{casa9}). The charge particle will experience the centripetal
force $(-m\omega^2 R\hat r_q)$ acting towards the center.
Now,
\begin{equation}
\vec s(t_{Ret}) = (x - R\cos \omega t_{Ret } )\hat i + (y - R\sin
\omega t_{Ret })\hat j + z\hat k
\end{equation}
The velocity and acceleration are
\begin{eqnarray}
  \dot{\vec r}_q(t_{Ret}) &=& -R\omega \sin \omega t_{Ret} \hat i +R\omega co{\mathop{\rm s}\nolimits} \omega t_{Ret} \hat j \\
 \vec a(t_{Ret}) &=& \ddot{\vec r}_q(t_{Ret}) = -R\omega ^2 \cos \omega t_{Ret} \hat i - R\omega ^2 \sin \omega t_{Ret} \hat j
\end{eqnarray}
 The acceleration $ \vec a(t_{Ret})$ (Please see the Appendix) in the limit  $s \to 0^+$ yields,
\begin{eqnarray}
 \vec a(t_{Ret}) &=& -R\omega ^2 \cos \omega \left( {t - \frac{s}{c}} \right)\hat i - R\omega ^2 \sin \omega \left( {t - \frac{s}{c}} \right)\hat j \nonumber \\
 % &=&R\omega ^2 \left[ {\cos \omega t\cos \left( {\frac{{\omega s}}{c}} \right)\hat i + \sin \omega t\sin \left( {\frac{{\omega s}}{c}} \right)\hat i} \right] \\
 % &+& R\omega ^2 \left[ {\sin \omega t\cos \left( {\frac{{\omega s}}{c}} \right)\hat j - \cos \omega t\sin \left( {\frac{{\omega s}}{c}} \right)\hat j} \right] \\
 % &=& R\omega ^2 \left[ {\cos \omega t\hat i + \sin \omega t\frac{{\omega s}}{c}\hat i + \sin \omega t\hat j - \cos \omega t\frac{{\omega s}}{c}\hat j} \right] \\
  &=& -R\omega ^2 \left[ {\hat r_q - \frac{{\omega s}}{c}\hat {\theta}_q } \right]
\end{eqnarray}
The self-force is given by
\begin{equation}  \vec F^{Circle}_{self}= \mathop
{Lim}\limits_{s \to 0^ +  } \frac{2}{3}\frac{q^2R}{{4\pi \varepsilon
_0 c^2 s}} \omega ^2 \hat r_q - \frac{2}{3}\frac{q^2R}{{4\pi
\varepsilon _0 c^3 }}\omega ^3 \hat {\theta}_q  \end{equation} The
self-force experienced by the charged particle picks up both the
tangential component which is responsible for the radiation reaction
as well as the radial component which displays the singular behavior
in the limit  $s \to 0^+$.
\section{Mass Renormalization}
The equation of motion for radiating charged particle is given as:
%\[ m \vec {\dot v} = {\vec F}_{Ext} + {\vec F}_{Rad}   \]
\begin{eqnarray}
 m_B \dot {\vec v} &=& \vec F_{Ext}  + \vec F_{Self} \nonumber\\%=\vec F_{Ext}  + \vec F^{Div}_{Self}+\vec F^{Rad}\nonumber \\
  &=& \vec F_{Ext}  - \left(\frac{2}{3}\frac{{q^2 }}{{4\pi  \varepsilon _0 c^2s }}\right)\dot{ \vec v}(t) +
  \frac{2}{3}\frac{{q^2 }}{{4\pi  \varepsilon _0 c^3 }}\ddot{ \vec v}(t)\nonumber \\
  &=& \vec F_{Ext}  - m_{Em} \dot {\vec v}(t) + \frac{2}{3}\frac{{q^2 }}{{4\pi  \varepsilon _0 c^3 }}\ddot
   {\vec v}(t)\label
{cc}
 \end{eqnarray}
 where $m_B$ corresponds to the mass of the charged particle that is not associated
 with the radiation reaction and is called bare mass. The bare mass $m_B$ refers
 to the physical phenomena at arbitrary short distance surrounding the point charge.
 The bare mass is not directly
 related to quantities that one measures at finite spatial length from the charged particle.
 However, $m_{Em}=\frac{2}{3}\frac{{q^2 }}{{4\pi  \varepsilon _0 c^2s }}$, defined
 as the electromagnetic mass, arises due to the presence of the electromagnetic
 field. The electromagnetic mass $m_{Em}$ is divergent for the point
 charge ($s\to 0^+$).
Now, we may rewrite equation (\ref{cc}) as
\begin{equation}
(m_B  + m_{Em} )\dot {\vec v}(t) = \vec F_{Ext}  +
\frac{2}{3}\frac{{q^2 }}{{4\pi  \varepsilon_0 c^3 }}\ddot {\vec
v}(t) \end{equation} In order to tame the divergence, the process of
renormalization is implemented as follows: Since a point charge
causes an infinite electromagnetic mass, we assume it to be
($+\infty + m_R$) so its bare mass must be postulated to be minus
infinite ($-\infty $) so as to render the observable physical
(renormalized) mass
\begin{equation} m_R = m_B  + m_{Em} = m_B + \frac{2}{3}\frac{{q^2 }}{{4\pi  \varepsilon _0 c^2s }}
=\textup{Finite}\end{equation}
 finite. This shift is known as mass renormalization. The bare mass and the electromagnetic
mass are themselves not physical observables.\\
In the case of charged particle executing simple harmonic motion,
the divergence piece of the self force $ m_{Em}x_0 \omega ^2 \sin
\omega t$ acts away from the equilibrium position. Whereas for the
charged particle moving in a circle, the divergent piece of the
self-force appears in the form of the centrifugal force $m_{Em}
\omega ^2 R \hat r_q$. To have a sensible theory, these infinities
are made to absorb via mass renormalization to obtain the physically
observable mass.
%%%%%%%%%%%%%%%%%%%%%%%%%%%%%%%%%
\section{Conclusion}
We derive the expression for the self-force for a charged particle
in arbitrary non-relativistic motion in rather neat and
sophisticated way than that presented by Boyer \cite{boy} in the
specific context of a charged particle in uniform circular motion.
We discuss illustrative examples pertaining to radiation reaction
and obtain explicitly the divergence pieces in the expressions of
their respective self-forces.We discuss the concept of mass
renormalization which implements the renormalization prescription as
to how to tame the divergence.
\begin{acknowledgments}
The author would like to thank the anonymous reviewers for their
helpful comments and suggestions.
\end{acknowledgments}
\appendix*   % Omit the * if there's more than one appendix.
\section{Calculation of $\vec a(t_{Ret})$ for a Charge Particle Moving
in a Circle} In the limit $s\to 0^+$, we have
\begin{eqnarray}
 \cos \left( {\omega t - \frac{{\omega s}}{c}} \right)= \cos \omega t\cos\frac{{\omega s}}{c} + \sin\frac{{\omega s}}{c}\sin \omega t\cong \cos \omega t + \frac{{\omega s}}{c}\sin \omega t \\
 \sin \left( {\omega t - \frac{{\omega s}}{c}} \right)=  \sin \omega t\cos\frac{{\omega s}}{c}-\cos \omega t\sin\frac{{\omega s}}{c} \cong \sin \omega t - \frac{{\omega s}}{c}\cos \omega
t\label{sine}
 \end{eqnarray}
 The unit vectors $\hat r_q(t_{Ret})$ and $\hat {\theta}_q (t_{Ret})$ in the limit $s\to 0^+$ may be
 expressed as
\begin{eqnarray}
 \hat r_q(t_{Ret} ) = \cos \left( {\omega t - \frac{{\omega s}}{c}} \right)\hat i + \sin \left( {\omega t -
 \frac{{\omega s}}{c}} \right)\hat j \cong \hat r_q(t) - \frac{{\omega s}}{c}\hat {\theta}_q (t) \\
 \hat {\theta}_q (t_{Ret} ) = -\sin \left( {\omega t - \frac{{\omega s}}{c}} \right)\hat i +
  \cos \left( {\omega t - \frac{{\omega s}}{c}} \right)\hat j \cong \frac{{\omega s}}{c}\hat r_q(t) + \hat {\theta}_q (t)
\end{eqnarray}
The velocity $\dot{\vec r}_q(t_{{\mathop{\rm Re}\nolimits} t} )$ is:
\begin{eqnarray}
 \dot{\vec r}_q(t_{{\mathop{\rm Re}\nolimits} t} ) &=&R\omega \hat {\theta}_q (t_{{\mathop{\rm Re}\nolimits} t} )
 = R\omega \hat {\theta}_q (t - \frac{s}{c})\nonumber \\
  &=& R\omega \left[ {\frac{{\omega s}}{c}\hat r_q(t) + \hat {\theta}_q (t)} \right] + O(s^2 )
 \end{eqnarray}
The acceleration $ \vec a(t_{{\mathop{\rm Re}\nolimits} t} )$ reads
\begin{eqnarray}
 \vec a(t_{{\mathop{\rm Re}\nolimits} t} ) &=& \ddot{\vec r}_q(t_{{\mathop{\rm Re}\nolimits} t} ) =
 -R\omega ^2 \hat r_q(t_{{\mathop{\rm Re}\nolimits} t} ) \nonumber\\
 &=&  -R\omega ^2 \left[ {\hat r_q(t) - \frac{{\omega s}}{c}\hat {\theta}_q (t)} \right] + O(s^2 )
 \end{eqnarray}

\end{document}